\documentstyle[epsf,12pt]{article}
\begin{document}
\begin{titlepage}
\begin{center}
\vspace*{3cm}

\begin{title}
\bold {\Huge 
The Bose-Einstein effect \\    
~\\
and the joint WW decay}
\end{title}
\vspace{2cm}

\begin{author}
\Large {K. FIA{\L}KOWSKI\footnote{e-mail address:
uffialko@thrisc.if.uj.edu.pl},  R. WIT }
\end{author}\\
\vspace{1cm}
{\sl Institute of Physics, Jagellonian University \\
30-059 Krak{\'o}w, ul.Reymonta 4, Poland}
\vspace{3cm}
\begin{abstract}
The influence of the Bose - Einstein interference effect on  
the joint WW hadronic decay is discussed. It is shown that the weight method 
incorporating this 
effect into Monte Carlo generators produces in a natural way an excess of 
average 
multiplicity as compared to the independent decay of two W bosons. The 
quantitative results 
for the average multiplicity and momentum distribution of charged pions, 
obtained with a 
simple parametrization of weights compatible with the observed shape of the "Bose - 
Einstein 
ratio", agree well with the existing data.
\end{abstract}
\end{center}

\vspace{1cm}

\noindent
May 1998 \\
hep-ph/9805476
\end{titlepage}

\section{Introduction}
\par
With the advent of LEP data on the $e^+e^- \rightarrow WW$ process one started 
to discuss in 
detail possible effects which could influence the final state obtained from 
double hadronic 
decay and break the simple factorization picture. The original motivation was 
the concern on 
the possibility to use these data for the precision measurements of $W$ 
mass, crucial for the 
tests of the standard model. Although the original suggestions of possible large 
$W$
mass shifts [1] seem to be rather exaggerated (see refs.[2,3], and references quoted 
therein), there are other possible 
observables 
which discriminate between the existing models of space - time development of the 
hadronization process.
\par
There are two proposed effects which can make the final state from double 
hadronic $WW$ 
decay significantly different from a simple superposition of two systems coming 
from the 
decays of two $W$ bosons. First, there is a Bose - Einstein interference (called often
 "HBT effect" [4], and here denoted by "BE effect") for 
pairs of identical pions coming from two different $W$-s. Second, there may 
be so-called "colour reconnection" processes, in which a colour neutral hadron is 
formed by coloured partons from two $W$-s.
\par
This note is devoted to the BE effect. We refer only occasionally to some results on the 
colour reconnection effect [1,2,5]. In 
Section II we discuss qualitatively the average multiplicity of $WW$ decay. In Section 
III we present quantitative results obtained when implementing the BE effect into 
Monte Carlo (MC) generators by our weight method [6]. We 
conclude with Section IV. 

\section{Multiplicity as a signal for the BE effect: \\a qualitative 
discussion}
\par 
The BE effect can make the final state of the joint $WW$ hadronic decay different from a
simple superposition of 
two systems coming from independent decays of two $W$-s. Its average 
multiplicity $\overline n_{WW}$ may be not just twice the average multiplicity from 
single $W$ decay $\overline n_W$. Moreover, one may predict  which 
one is bigger.
\par
For the $WW$ production at the energies near to the threshold the decay products 
of both $W$-s are formed in the same space - time region. Therefore the 
transition amplitudes to the hadronic final states are not just products of two 
decay amplitudes.  
The symmetrization enhances the probabilities to obtain final states where the momenta of 
identical pions are close in momentum space. This is more likely for higher 
multiplicities, 
where many pions are slow in the CM frame. Thus one may expect that $\Delta \overline n 
\equiv 
\overline n_{WW} - 2\overline n_W > 0$. 
Moreover, the excess should be located at low $CM$ momenta, since the decay products of 
different $W$-s with high $CM$ momenta are rarely close in momentum space.
\par
This argument is qualitative and in general one cannot prove that the excess is 
significant. 
Nevertheless, it seems strange to claim that "there is no reason to expect the excess of 
multiplicity from the BE effect"[5]. There exists a specific algorithm imitating the 
BE effect in MC generators  which does not change the 
multiplicities [7,8,9], but it has no theoretical justification. The excess of 
multiplicity 
in joint 
$WW$ decay seems to be a natural result of the BE effects.
\par
This should be contrasted with the consequences of the colour reconnection effects, which 
lead 
naturally to the deficit of multiplicity [5] due to the colour screening. Thus the net 
result of these two effects may well be negligible (the so called "BE conspiracy" [9]). 
However, 
the data may also show that one 
of the effects dominates. For this reason it is useful to formulate the quantitative 
predictions. It will be done for the BE effect in the next Section within the 
framework of the weight method, for which one can calculate easily the multiplicity 
excess.
\par
 Let us 
remind here shortly the basics of the weight method [10]. The probability to produce any 
final 
state corrected for the BE effect may be approximated as a product of the original 
probability (without the BE effects) and the weight $W(n)$ calculated as a sum of $n!$ 
terms, 
each 
being a product of two - particle weight factors:
\begin{equation}
\label{s1}
W(n) = \sum_{\{P_n\}}\prod_{i=1}^{n}w_{i,k_i}.
\end{equation}                                                                         
\par
More precisely, the global BE weight is a product of weights calculated for each kind of 
identical particles. The shape of a two- particle weight factor should be fitted to the 
data. 
A simple one-parameter Gaussian form is often used :
\begin{equation}
\label{s3}w_{i,k}(p_i,p_k)=exp[-(p_i-p_k)^2/2\sigma^2].
\end{equation}
\par
The weight method meets obvious practical difficulties for a large number of particles 
$n$, 
as the factorial increase of number of terms makes the computing impossibly long. There 
are 
some methods to overcome this difficulty [11,12,13,14,15]. We employ here the recent 
version of our 
method [6,3], which uses the clustering algorithm with the cluster size defined by a 
parameter $\epsilon$, and yields the weight as a product of cluster 
weight factors. For clusters with more than five particles these factors are calculated 
approximately. The employed method is  time saving and we regard it as a more reliable 
approximation of formula (1) than the previous ones.
\par
We are not going to describe here this method in details. Let us remind only two 
important 
approximations used. First, as in the earlier papers [7,12,13,15], only so called "direct 
pions" 
are taken into account since the decay products of the long - living resonances are born 
far 
away 
from the collision area and the parameter $\sigma$ of formula (2) would be very small for 
pairs including such particles. Second, a simple rescaling of weights by a factor $cV^n$ 
[12] 
is used instead of refitting all the free parameters of the MC code (fitted to the data 
without the BE effect). This allows to restore the average multiplicity increased by the 
weights, which are naturally larger for large multiplicities.
\par
It is this second approximation which allows to estimate easily the multiplicity excess 
due 
to the BE 
effect. For the MC without the BE effect the final state of the hadronic $WW$ decay is a 
simple superposition of two $W$ decay product systems. This means that the generating 
function of the multiplicity distribution
\begin{equation}
\label{s7}
G(z) = \sum P(n)z^n.
\end{equation} 
 should be just the square of the generating 
function for a single decay
\begin{equation}
\label{s4}
G_{WW}(z) = [G_W(z)]^2.
\end{equation} 
\par 
Rescaling the distribution $P(n)$ by  $cV^n$ factors rescales the argument of $G$ by $V$ 
(the 
normalization factor $c$ is irrelevant since any $G(z)$ has to fulfill the equation $G(1) 
= 
1$). 
This does not spoil the relation (4). Rescaling may be interpreted as refitting the 
parameter 
which controls the density of particles from a single string. Thus the same value of the 
rescaling parameter V should be used for single- and double $W$ decay.       
\par
In the single decay this value is fitted to restore exactly the average multiplicity 
obtained 
without weights (and compatible with the data). However, for the double decay such a 
reduction is insufficient: the BE weight for a final state from the double decay is 
always 
bigger than the product of weights  for two independent decays. Since weights enhance the 
states with high multiplicity, an excess of multiplicity appears. It will be presented 
quantitatively in the next section.
\par
Let us conclude this discussion with a remark on the energy dependence. For the energies 
far 
above the threshold both $W$-s move apart with nearly the speed of light and the 
formation 
points 
of their hadronic decay products are separated by sizeable distances. Thus both the BE 
effect 
and the colour reconnection effect should be greatly reduced and the factorization of the 
final state should be a very goood approximation. Later on we comment shortly on the 
possible 
parametrization of the energy dependence of the BE effect.

\section{Results and comparison with data}
\par
We have generated many samples of 100 000 events for different channels of the $e^+e^- 
\rightarrow 
W^+W^-$ process at 172 GeV by the default PYTHIA/JETSET generator. First, we have chosen 
the 
channel 
where one of the $W$-s decays leptonically, and the other one decays into two hadronic 
jets. 
We have 
generated a sample of events without weights and samples with weights given by our 
approximation to the 
formula (1) (with the two-particle weight factors given by (2) and various values of 
$\sigma$ and $\epsilon$). Then we have chosen the values of the 
parameters
$\sigma =  0.2$ GeV and the ratio $\epsilon/\sigma^2 = 10$ (at the lower limit of the 
range 
of 
values, for which the results do not depend on this ratio [6]). With these values we have 
found the 
average multiplicity with BE weights by 20\%  larger than without weights. To restore the 
original 
multiplicity of charged pions we rescaled the weights by a $cV^n$ factor with V = 0.920 
and 
c = 
1.798. This last value guaranteed that  the average value of the rescaled weight is just 
1. 
Rescaling reduces the discrepancy between the average multiplicities to less than 1\%. 
Also the inclusive 
distributions are now not significantly affected by the BE weights.

\par
The choice of the value of $\sigma$ was determined by the experimental shape of BE 
enhancement 
at the 
$Z^0$ peak (where the data are much better than for the $W$ decay, and the shape is 
expected 
to be 
similar). For simplicity we use the same quantity as recently employed for the BE 
analysis of 
the LEP2 data [16]: the "double ratio", defined as 
\begin{equation}
\label{s5}
R^{++}_{+-}(Q) = (\rho^{++}_2/\rho^{+-}_2)^{exp}/(\rho^{++}_2/\rho^{+-}_2)^{MC},
\end{equation} 
\noindent
where $\rho_2^{ik}$ are the two-particle densities in $Q = \sqrt{-(p_1-p_2)^2}$ for the 
pairs 
of 
pions with 
different charge combinations. The superscripts "exp" and "MC" denote the data and a MC, 
in 
which 
the BE effect is not included. In Fig.1 we show the data from ref. [16] and the 
analoguous 
"double 
ratio" of the MC results with weights (rescaled) to those without weights. It is obvious 
that our 
weight method with the chosen value of $\sigma$ describes the data well.

\vspace{0.5cm}
\epsfxsize=10cm
~~~~~~~~~~~~~\epsfbox{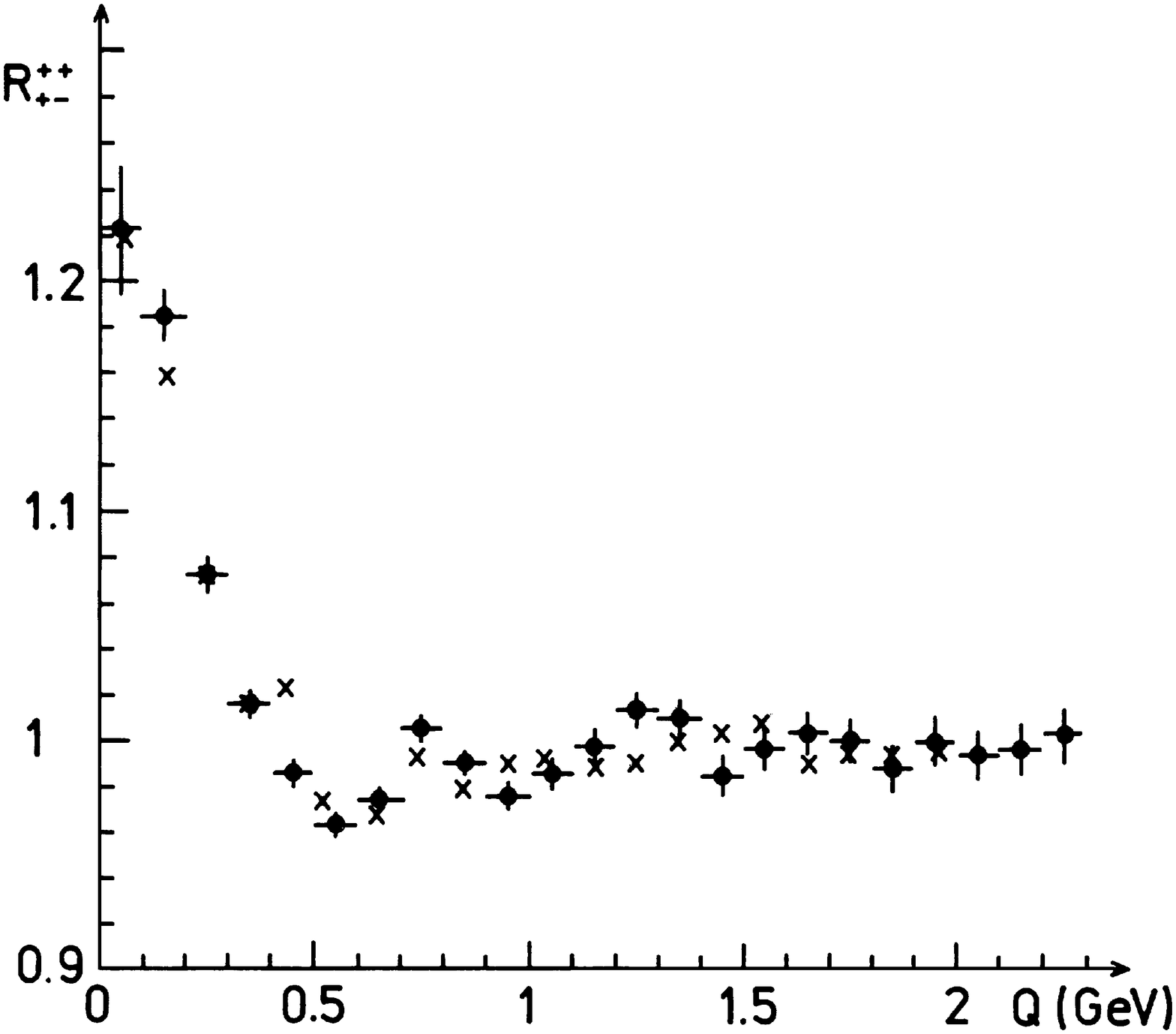}
\vspace{0.5cm}
\par
 {\bf Fig.1.} {\sl The double ratio (4) from the data of Ref. [15] (black points) and 
from 
our 
MC (crosses). The errors for MC, not shown for transparency, are of the order of 0.01.}  
\vspace{0.5cm}
\par
Next, we generate the samples of events in which both $W$-s decay into two hadronic jets. 
As 
argued in the 
previous section, the rescaling of weights should be done with the same value of the 
parameter 
$V$ 
as for a single $W$ hadronic two-jet decay. For simplicity we use the same value $\sigma 
= 
0.2$ GeV 
for all pairs of identical pions, as the energy is not far above the threshold (as we 
argue 
later, 
the effective value of $\sigma$ for pairs of pions from different $W$-s should decrease 
with 
energy). The value 
of $c$ is again adjusted to give the average weight equal one. To the seven events with  
anomalous high weight values we ascribed a limiting value of 500 to avoid excessive 
fluctuations in some histograms. The histograms shown in this paper are not affected by 
this cut.
\par
In Fig.2 we present the distributions of CM momenta for charged pions with- and without 
weights. 
One 
sees that the sample with weights shows a clear excess of pions for momenta below 1.5 
GeV. 
The 
global 
excess $\Delta \overline n$ integrated over momenta is about 2.1 charged pions per event. 
Since 
without 
weights 
we find the average multiplicity just twice the multiplicity of a single $W$ decay, this 
excess 
should be atributed to the BE effect. Let us note that the results are quite sensitive to 
the 
value 
of $\sigma$: reducing it by 30\% we get a four times smaller excess! We estimate the 
uncertainty of 
the value of $\sigma$ to be about 10\%, resulting in the uncertainty of $\Delta \overline 
n$ of about 
40\%. The obtained value of $\Delta \overline n = 2.1 \pm 0.9$ is in a very good 
agreement with the 
averaged preliminary data from 
all the LEP2 experiments [17], which give $\Delta \overline n = 2.4 \pm 1.8$. Obviously, 
much better 
data are 
needed to draw any definite conclusions, but the results look encouraging.
\par
\vspace{0.5cm}
\epsfxsize=10cm

~~~~~~~~~~~~~\epsfbox{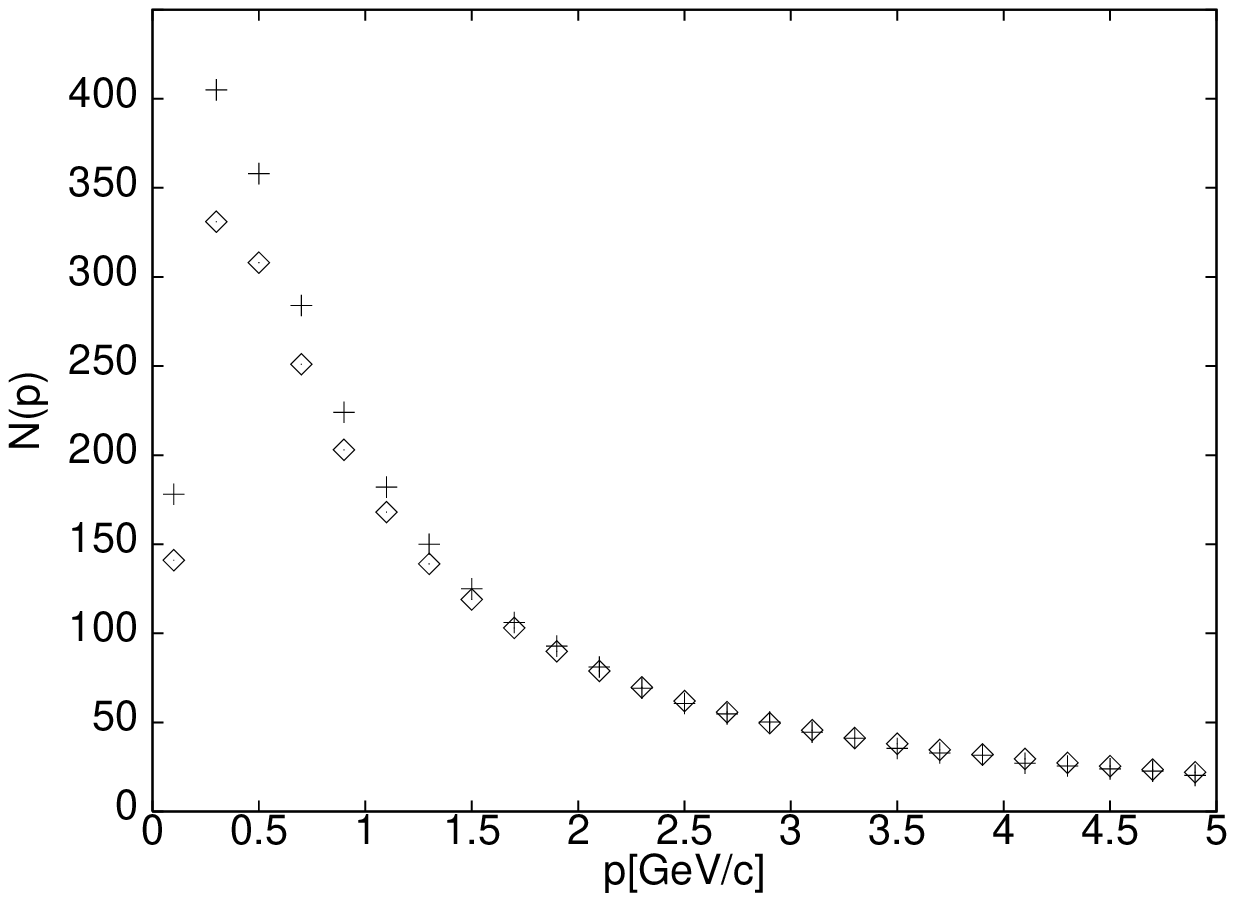}

\vspace{1cm}
\par
 {\bf Fig.2.} {\sl The distribution  of CM momenta $p[GeV/c]$ for the $e^+e^- \rightarrow 
W^+W^- \rightarrow 4j$ (in thousands of pions) without weights (diamonds)
and with rescaled weights (crosses).}   \\
\par
 The fact that the excess 
appears at small momenta agrees also with some preliminary data [17]. Let us note that 
our 
estimate of $\Delta \overline n$ 
is much larger than the estimates of the (negative) shift of the multiplicity due to the 
colour 
reconnection (CR) effect [5]. This justifies {\it a posteriori} our procedure of 
calculating 
the BE 
effect neglecting the possible CR effects. If the future results confirm the positive 
shift 
of 
multiplicity of similar size as seen now, it will be a strong argument for our method of 
implementing the BE effect, and against an important r\^ole of the CR effects.

\section{Conclusions and outlook}
\par
As a new application of our method of implementing the BE effect in MC generators [6], we 
have  
calculated the excess of the charged pion multiplicity in the joint hadronic $WW$ decay 
over 
the 
value expected from independent decays of two $W$-s. The value of this excess is 
significantly 
larger than the deficit expected from the colour reconnection effects [5], and agrees 
very 
well with 
the preliminary data [17]. The excess is located at small CM momenta.
\par
If the preliminary data are confirmed, it will be a decisive argument for using the 
weight 
method 
and not the momentum shifting methods [7,8,9] for implementing the BE effects in MC 
generators. In 
fact, any observed excess leads to the same conclusion: the momentum shifting method does 
not 
affect 
the multiplicity and the colour reconnection effects produce a deficit instead of an 
excess.
\par
Both effects mentioned above are expected to decrease at higher energy. For the BE effect 
a 
simple 
approximate method to estimate this decrease would be to define an effective source size 
$R_{eff} = 
hc/\sigma_{eff} $ for pairs of pions from different $W$-s as given by 
\begin{equation}
\label{s6}
R^{2}_{eff} = R^{2}_{W} + 4\beta^2 \gamma^2 c^2 \tau^2
\end{equation} 
where $\beta$ and $\gamma$ are the velocity and the Lorentz factor of  each $W$ in the CM 
frame 
at given energy. $R_W$, given by $hc/\sigma$, defines the effective size of $W$, as 
measured 
by the 
BE effect in the $W$ (or $Z$) decay, and should be used for pairs coming from the same 
$W$.  
We do 
not present here any definite predictions for the energy dependence, as it is not quite 
clear 
if 
$\tau $ should represent $W$'s lifetime or rather the hadronization time for $W$ decay 
products; this 
problem should be solved experimentally.
\par
A more direct way to investigate the BE effect for the joint $WW$ decay is to 
subtract the distributions for the joint $WW$ decay and for the single $W$ decay to get 
the 
separated BE 
effect for pairs from two different $W$-s. However, this  has not given yet conclusive 
results [16]. Some 
preliminary data suggest that at 172 Gev the value of $R_{eff}$ for such pairs should be 
similar to 
that of $R$, whereas other data show a supression instead of BE enhancement at low Q (!). 
Certainly 
better data are needed to clarify the situation. In any case the 
conclusions 
drawn 
from the multiplicity excess/deficit and from the low $Q$ enhancement/suppresion should 
be 
compatible.

\vspace{0.5cm}
{\Large \bf Acknowledgements}
\vspace{0.2cm}
\par
We thank A. Bia{\l}as and K. Zalewski for reading the manuscript. A financial  support 
from KBN grants No 2 P03B 086 14 and No 2 P03B 044 12
is gratefully acknowledged. 
\vspace{1cm}

{\Large \bf References}
\vspace{0.2cm}

\noindent
[1] J. Ellis and K. Geiger, {\it Phys.Rev.} {\bf D54} (1996) 1967.

\noindent
[2] B.R. Webber, {\it J. Phys.} {\bf G24} (1998) 287.

\noindent
[3] K. Fia{\l}kowski and R. Wit, e-print hep-ph/9805223, talk presented at the XXXIII-rd
Moriond Meeting, Les Arcs 1998, to be published in the Conference Proceedings. 

\noindent 
[4] R. Hanbury-Brown and R.Q. Twiss,  {\it Nature} {\bf178}
(1956) 1046.

\noindent
[5] V.A. Khoze and T. Sj\"ostrand, e-print hep-ph/9804202. 

\noindent
[6] K. Fia{\l}kowski, R. Wit and J. Wosiek, e-print hep-ph/9803399.

\noindent
[7] T. Sj\"ostrand and M. Bengtsson, {\it Comp. Phys.
 Comm.} {\bf 43} (1987) 367; T. Sj\"ostrand, {\it Comp. Phys. Comm.}
{\bf 82} (1994) 84.

\noindent
[8] L. L\"onnblad and T. Sj\"ostrand, {\it Phys.Lett.} {\bf B351}
(1995) 293.

\noindent
[9] L. L\"onnblad and T. Sj\"ostrand, {\it Eur.
Phys. J.} {\bf C2} (1998) 165.

\noindent
[10] A. Bia{\l}as and A. Krzywicki, {\it Phys.Lett.} {\bf B354} (1995) 134.

\noindent
[11] S. Haywood, Rutherford Lab. Report RAL-94-074.

\noindent
[12] S. Jadach and K. Zalewski,  {\it Acta Phys. Pol.}
{\bf B28} (1997) 1363.

\noindent
[13] K. Fia{\l}kowski and R. Wit,  {\it Acta Phys. Pol.}
{\bf B28} (1997) 2039.

\noindent
[14] V. Kartvelishvili, R. Kvatadze and R. M{\o}ller,
e-print hep-ph/9704424.

\noindent
[15] K. Fia{\l}kowski and R. Wit, {\it Eur. Phys. J.} {\bf C2} (1998) 691.

\noindent
[16] H.R. Hoorani, talk presented at the XXXIII-rd
Moriond Meeting, Les Arcs 1998, \\to be published in the Conference Proceedings. 

\noindent
[17] A. De Angelis, talk presented at the XXVII-th Int. Symp. on Multiparticle
Dynamics, Frascati 1997, to be published in the Symposium Proceedings,
{\it Nucl. Phys.} C, 1998; \\N.  Pukhayeva, ibid.

\end{document}